\documentclass[a4paper,fleqn,usenatbib]{mnras}
\usepackage[T1]{fontenc}
\usepackage{ae,aecompl}
\usepackage{graphicx}
\usepackage{amsmath}
\usepackage{amssymb}
\usepackage{txfonts}

\def\spirou#1{{#1}}
\def\foca#1{{ #1}}

\newcommand{\ltsima}{$\; \buildrel < \over \sim \;$}
\newcommand{\lsim}{\lower.5ex\hbox{\ltsima}}
\newcommand{\gtsima}{$\; \buildrel > \over \sim \;$}
\newcommand{\gsim}{\lower.5ex\hbox{\gtsima}}
\newcommand{\bra}{\langle}
\newcommand{\ket}{\rangle}

\newcommand{\dd}{\mathrm{d}}

\newcommand{\ci}{\mathrm{i}}

\newcommand{\chip}{{\chi^\prime}}

\newcommand{\likeli}{\mathcal{L}}

\title[Fisher-matrix across parameter space]
{Describing variations of the Fisher-matrix across parameter space}
\author[B.M. Sch{\"a}fer, R. Reischke]
{Bj{\"o}rn Malte Sch{\"a}fer\thanks{e-mail: bjoern.malte.schaefer@uni-heidelberg.de}, Robert Reischke\\
Astronomisches Recheninstitut, Zentrum f{\"u}r Astronomie der Universit{\"a}t Heidelberg, Philosophenweg 12, 69120 Heidelberg, Germany}

\begin{document}
\pagerange{\pageref{firstpage}--\pageref{lastpage}}
\pubyear{2016}
\maketitle
\label{firstpage}

\begin{abstract}
Forecasts in cosmology, both with Monte-Carlo Markov-chain methods and with the Fisher matrix formalism, depend on the choice of the fiducial model because both the signal strength of any observable as well as the model nonlinearities linking observables to cosmological parameters vary in the general case. In this paper we propose a method for extrapolating Fisher-forecasts across the space of cosmological parameters by constructing a suitable basis. We demonstrate the validity of our method with constraints on a standard dark energy model extrapolated from a $\Lambda$CDM-model, as can be expected from 2-bin weak lensing tomography with a Euclid-like survey, in the parameter pairs $(\Omega_m,\sigma_8)$, $(\Omega_m,w_0)$ and $(w_0,w_a)$. Our numerical results include very accurate extrapolations across a wide range of cosmological parameters in terms of shape, size and orientation of the parameter likelihood, and a decomposition of the change of the likelihood contours into modes, which are straightforward to interpret in a geometrical way. We find that in particular the variation of the dark energy figure of merit is well captured by our formalism.
\end{abstract}


\section{Introduction}
Forecasting techniques based on the Fisher-formalism are ubiquitous in cosmology: They are straightforward to implement, they are commonly used to investigate to what extent cosmological models can be tested with surveys, and they are employed to optimise survey design and to maximise sensitivity on certain cosmological parameters. They have been applied to forecasts from the cosmic microwave background \citep{perotto_probing_2006, elsner_fast_2012} including cross correlations \citep{2014PhRvD..89d3516P}, tomographic weak lensing \citep{,1999ApJ...522L..21H,2002PhRvD..66h3515H,2003PhRvL..91n1302J,2004MNRAS.348..897T, 2004ApJ...601L...1T, 2005PhRvD..72d3002H, 2007MNRAS.381.1018A, 2012MNRAS.423.3445S}, 3d weak cosmic shear \citep{2003MNRAS.343.1327H, kitching_3d_2014,grassi_detecting_2014}, weak gravitational lensing of the cosmic microwave background \citep{2006JCAP...10..013P, PhysRevD.69.083514, 2013PhRvD..88d3505S}, other secondary anisotropies including the Sunyaev-Zel'dovich and integrated Sachs-Wolfe effects, cluster cosmology \citep{2003PhRvD..67h1304H, 2009ApJ...698L..33M, 2010PhRvD..82h1301K, 2010PhRvD..82d1301K, 2011ARA&A..49..409A, 2012MNRAS.422...44P, 2013JCAP...02..030K, 2015arXiv150502165S}, galaxy clustering including baryon acoustic oscillations \citep{2009MNRAS.397.1348W, seisen}, supernova cosmology \citep{efstathiou_constraining_1999} and many more.

At the core of the Fisher-formalism \citep{1997ApJ...480...22T, coe_fisher_2009, 2009arXiv0906.0974B, 2011IJMPD..20.2559B} is the approximation of the parameter likelihood with a Gaussian function. Forecasting of errors and degeneracies needs to assume a fiducial model, which is set to be yield the maximal likelihood, and one proceeds to compute the curvature of the logarithmic likelihood, which after averaging over all possible realisations of the data, quantifies the inverse errors on the parameters one intends to measure. The results of the Fisher-formalism hinge on the choice of a fiducial model, which sets the signal strength and determines how nonlinearities in the model are approximated. Both effects lead to variations of the Fisher matrix across parameter space which is an important issue as some choices of parameters can exclude certain physical effects, for instance dark energy perturbations cease as the equation of state $w$ approaches $w=-1$.

\spirou{In this paper we intend to investigate the dependence of Fisher-forecasts on the choice of the reference cosmology and propose a method to describe and quantify variations of the Fisher-matrix across parameter space: For this purpose, we construct a transformation between Fisher-matrices that are infinitesimally displaced along the axes of parameter space and show that these infinitesimal transformation can be assembled to arbitrary displacements. This not only allows us to predict the rate of change of the Fisher matrix with cosmological parameters but also to quantify the type of change in terms of changes in size,  orientation and shape of the Fisher-ellipse. Apart from this theoretical result the formalism is suited to be applied to optimisation of survey design or for speeding up Monte-Carlo Markov-chains by updating the proposal distribution for sampling.}

As an application we consider constraints on the cosmological parameters $\Omega_m$, $\sigma_8$, $w_0$ and $w_a$ from 2-bin weak lensing tomography with Euclid\footnote{http://www.euclid-ec.org/}, but our method can in principle be applied to any Gaussian forecasting problem in cosmology. Lensing tomography makes the best use of weak lensing data by separating the influence of cosmological parameters on the signal at different redshifts. Euclid's weak lensing signal will be quite strong: Integrated over all multipoles including nonlinear scales and in a full tomographic setup with many redshift bins the signal will have a significance of close to $1000\sigma$, allowing sub-percent precision on cosmological parameters.

The fiducial cosmological model is a spatially flat $\Lambda$CDM-cosmology, with specific parameter choices $\Omega_m = 0.25$, $n_s = 1$, $\sigma_8 = 0.8$ and $h=0.7$, in comparison to generic quintessence models whose equation of state is parameterised by a linear time-evolution in $w_0$ and $w_a$, in contrast to the fiducial cosmology, where $w_0=-1$ and $w_a=0$. Throughout the paper we assume the summation convention implying automatic summation over repeated indices.

After a summary of cosmology in Sect.~\ref{sect_cosmology} and weak gravitational lensing in Sect.~\ref{sect_lensing} we outline the statistical methods in Sects.~\ref{sect_statistics} and ~\ref{sect_lie}. We summarise and discuss our results in Sect.~\ref{sect_summary}.

\section{cosmology}\label{sect_cosmology}
Under the symmetry assumption of Friedmann-Lema{\^i}tre-cosmologies all fluids are characterised by their density and their equation of state: In spatially flat cosmologies with the matter density parameter $\Omega_m$ and the corresponding dark energy density $1-\Omega_m$ one obtains for the Hubble function $H(a)=\dot{a}/a$ the expression,
\begin{equation}
\frac{H^2(a)}{H_0^2} = \frac{\Omega_m}{a^{3}} + \frac{1-\Omega_m}{a^{3(1+w_0+w_a)}}\times\exp(-3w_aa),
\end{equation}
where a linearly evolving equation of state function $w(a)$ \citep{2001IJMPD..10..213C, 2006APh....26..102L, 2008GReGr..40..329L},
\begin{equation}
w(a) = w_0 + (1-a)w_a,
\end{equation}
was assumed for the dark energy component. The comoving distance $\chi$ is related to the scale factor $a$ through
\begin{equation}
\chi = -c\int_1^a\:\frac{\dd a}{a^2 H(a)},
\end{equation}
where the Hubble distance $\chi_H=c/H_0$ sets the distance scale for cosmological distance measures. Cosmic deceleration $q=\ddot{a}a/\dot{a}^2$ is related to the logarithmic derivative of the Hubble function, $2-q = 3+\dd\ln H/\dd\ln a$.

Small fluctuations $\delta$ in the distribution of dark matter grow, as long as they are in the linear regime $\left|\delta\right|\ll 1$, according to the growth function $D_+(a)$ \citep{2003MNRAS.346..573L,1998ApJ...508..483W},
\begin{equation}
\frac{\dd^2}{\dd a^2}D_+(a) +
\frac{2-q}{a}\frac{\dd}{\dd a}D_+(a) -
\frac{3}{2a^2}\Omega_m(a) D_+(a) = 0,
\label{eqn_growth}
\end{equation}
and their statistics is characterised by the spectrum $\bra \delta(\bmath{k})\delta^*(\bmath{k}^\prime)\ket = (2\pi)^3\delta_D(\bmath{k}-\bmath{k}^\prime)P_\delta(k)$. Inflation generates a spectrum of the form $P_\delta(k)\propto k^{n_s}T^2(k)$ with the transfer function $T(k)$ \citep{eisenstein_power_1999, eisenstein_baryonic_1998} which is normalised to the variance $\sigma_8$ smoothed to the scale of $8~\mathrm{Mpc}/h$,
\begin{equation}
\sigma_8^2 = \int_0^\infty\frac{k^2\dd k}{2\pi^2}\: W^2(8~\mathrm{Mpc}/h\times k)\:P_\delta(k),
\end{equation}
with a Fourier-transformed spherical top-hat $W(x) = 3j_1(x)/x$ as the filter function. From the CDM-spectrum of the density perturbation the spectrum of the dimensionless Newtonian gravitational potential $\Phi$ can be obtained,
\begin{equation}
P_\Phi(k) \propto \left(\frac{3\Omega_m}{2\chi_H^2}\right)^2\:k^{n_s-4}\:T(k)^2,
\end{equation}
by applying the comoving Poisson-equation $\Delta\Phi = 3\Omega_m/(2\chi_H^2)\delta$ for deriving the gravitational potential $\Phi$ from the density $\delta$. Nonlinear structures increase the variance on small scales, which is described through a parameterisation of $P_\delta(k,a)$ proposed by \citet{2003MNRAS.341.1311S}.

\section{weak gravitational lensing}\label{sect_lensing}
In weak gravitational lensing one investigates the action of gravitational tidal fields on the shape of distant galaxies by the distortion of light bundles \citep[for reviews, please refer to][]{2001PhR...340..291B, hoekstra_weak_2008, huterer_weak_2010, 2010CQGra..27w3001B}. The lensing potential $\psi_i$ is given by a projection integral,
\begin{equation}
\psi_i = \int_0^{\chi_H}\dd\chi\:W_i(\chi)\Phi,
\label{eqn_lensing_potential}
\end{equation}
relating $\psi_i$ to the gravitational potential $\Phi$ through weighting function $W_i(\chi)$,
\begin{equation}
W_i(\chi) = 2\frac{D_+(a)}{a}\frac{G_i(\chi)}{\chi}.
\end{equation}
As a line of sight-integrated quantity, the projected potential contains less information than the sourcing field $\Phi$. In order to partially regain that information, one commonly divides the sample of lensed galaxies into $n_\mathrm{bin}$ redshift bins and computes the lensing signal for each of the bins $i$ separately. Therefore, one defines the tomographic lensing efficiency function $G_i(\chi)$,
\begin{equation}
G_i(\chi) = \int^{\chi_{i+1}}_{\mathrm{min}(\chi,\chi_i)}\dd\chip\: 
p(\chip) \frac{\dd z}{\dd\chip}\left(1-\frac{\chi}{\chip}\right),
\end{equation}
with $\dd z/\dd\chip = H(\chip) / c$ and the bin edges $\chi_i$ and $\chi_{i+1}$, respectively. Euclid forecasts use the parameterisation of the redshift distribution $p(z)\dd z$,
\begin{equation}
p(z)\dd z \propto \left(\frac{z}{z_0}\right)^2\exp\left[-\left(\frac{z}{z_0}\right)^\beta\right]\dd z,
\end{equation}
\spirou{with $\beta=3/2$ causing a slightly faster than exponential decrease at large redshifts \citep{EuclidStudyReport}}.

Combining all results one obtains the angular spectra $C_{\psi,ij}(\ell)$ of the tomographic weak lensing potential $\psi_i$ in the flat-sky approximation \citep{1954ApJ...119..655L},
\begin{equation}
C_{\psi,ij}(\ell) = \int_0^{\chi_H}\frac{\dd\chi}{\chi^2}\:W_i(\chi)W_j(\chi)\:P_\Phi(k=\ell/\chi).
\end{equation}
Weak lensing convergence is related to the lensing potential by applying the Laplace operator, therefore its spectra is equal to $\ell^4C_{\psi,ij}(\ell)/4$. Observed spectra of the weak lensing shear will contain a constant shape noise contribution $\sigma_\epsilon^2 n_\mathrm{bin}/\bar{n}$,
\begin{equation}
\hat{C}_{\psi,ij}(\ell) = C_{\psi,ij}(\ell) + \sigma_\epsilon^2\frac{n_\mathrm{bin}}{\bar{n}}\ell^4\times\delta_{ij}.
\end{equation}
The spectra $C_{\psi,ij}(\ell)$ is not zero for $i\neq j$ leading to a non-diagonal covariance matrix in the 
Fisher-matrix construction. We chose the bin edges in a way that they contain identical fractions of the total number $4\pi f_\mathrm{sky}\bar{n}$ of galaxies, with the number $\bar{n}$ of galaxies per unit solid angle and the sky coverage of the survey $f_\mathrm{sky}$.

\section{statistics}\label{sect_statistics}
\foca{As an application of our formalism we will consider a tomographic weak lensing survey such as Euclid's, which are able to deliver competitive bounds on the density parameters $\Omega_m$ and $\Omega_w$, the dark energy equation of state parameters $w_0$ and $w_a$ and the fluctuation amplitude $\sigma_8$. We should emphasise that our method of describing Fisher-matrix variations across parameter space is not restricted to weak lensing or to any specific construction of the Fisher-matrix, because we only require for our formalism that the matrix in question in positive definite, which is always applicable to the negative second derivative of the logarithmic likelihood.}

Full sky tomographic weak lensing surveys provide a measurement of $2\ell+1$ statistically independent modes for each multipole $\ell$, and independent multipoles in the case of homogeneous and isotropic random fields, from which constraints on cosmological parameters can be derived \citep{1997ApJ...480...22T, 2002PhRvD..66h3515H}. Technically, one would derive the set of modes $\psi_{\ell m,i}$ of modes separately for each tomography bin by spherical harmonical transform,
\begin{equation}
\psi_{\ell m,i} = \int\dd\Omega\: \psi_i(\bmath\theta)Y_{\ell m}^*(\bmath\theta).
\end{equation}
The likelihood that a model $\hat{C}_\psi(\ell)$ is able to reproduce the set $\left\{\psi_{\ell m,i}\right\}$ of observed modes $\psi_{\ell m,i}$ separates in $\ell$ and $m$ according to
\begin{equation}
\likeli\left(\left\{\psi_{\ell m,i}\right\}\right) = \prod_\ell \likeli\left(\psi_{\ell m,i}|\hat{C}_{\psi,ij}(\ell)\right)^{2\ell + 1},
\end{equation}
because of the symmetry assumptions, while there is no separation in the tomographic bin index $i$,
\begin{equation}
\likeli\left(\psi_{\ell m,i}\right) = 
\frac{1}{\sqrt{(2\pi)^{n_\mathrm{bin}}\mathrm{det}\hat{C}_\psi(\ell)}}\exp\left(-\frac{1}{2}\psi_{\ell m,i}(\hat{C}_{\psi}(\ell)^{-1})_{ij}\psi_{\ell m,j}\right),
\end{equation}
from the fact that both the cosmic structure as well as the noise are Gaussian random fields. This assumption is limited to linear structure formation, where all Fourier-modes evolve in a statistically independent way. Despite the fact that the line of sight-integrations reduce the amount of non-Gaussianity in the lensing observable due to the central limit theorem, non-Gaussian covariances lead to misestimations of parameters, as shown by \citet{scoccimarro_power_1999, kayo_information_2013}. Our model for nonlinear structures effectively increases the variance of the fields without accounting for the deviation from Gaussianity.

The logarithmic likelihood $L = -\ln\likeli$ is equal to 
\begin{equation}
L = \sum_\ell\frac{2\ell+1}{2}\:\left(\ln\hat{C}_{\psi,ii} + (\hat{C}_\psi^{-1})_{ij}\:\psi_{\ell m,i}\psi_{\ell m,j}\right),
\end{equation}
up to an additive constant. From the data-averaged curvature $F_{\mu\nu} = \bra\partial_\mu\partial_\nu L\ket$ of the negative logarithmic likelihood one derives the Fisher matrix $F_{\mu\nu}$,
\begin{equation}
F_{\mu\nu} = \sum_\ell\frac{2\ell+1}{2}\left(
\partial_\mu\ln\hat{C}_{\psi,ij}(\ell)\:\partial_\nu\ln\hat{C}_{\psi,ji}(\ell)\right),
\label{eqn_fisher}
\end{equation}
with $\partial_\mu$ being derivatives with respect to individual cosmological parameters, $x_\mu = \left(\Omega_m,\sigma_8,w_0,w_a\right)$, \spirou{for the case of vanishing expectation values of individual modes, $\bra\psi_{\ell m,i}\ket = 0$.} \foca{This, however, is not required for the construction of the Lie-basis, as the only requirement is the positive definiteness of the object under consideration, which $F_{\mu\nu}$ fulfills by definition even in the most general case.}  

The Fisher matrix $F_{\mu\nu}$ allows the definition of an on average expected Gaussian posterior distribution of inferred parameters,
\begin{equation}
p(x_\mu) = \frac{1}{\sqrt{(2\pi)^n\det F^{-1}}}\exp\left(-\frac{1}{2}\Delta x_\mu F_{\mu\nu} \Delta x_\nu\right),
\end{equation}
with the dimension $n$ of the parameter space and the distances $\Delta x_\mu$ of the parameters from their fiducial choice.

The characteristics of Euclid's weak lensing survey \citep{2008arXiv0802.2522R, EuclidStudyReport} are summarised as $(i)$ a median redshift of $0.9$, $(ii)$ a yield of $\bar{n}=4.7\times10^8$ galaxies per unit solid angle, $(iii)$ a sky fraction of $f_\mathrm{sky} \simeq 0.363$ and $(iv)$ a Gaussian shape noise with standard deviation $\sigma_\epsilon=0.3$. For demonstrating the construction of a Lie-basis we will work with $n_\mathrm{bin}=2$ tomographic bins and collect the signal up to the multipole $\ell=1000$. We work with modes $\psi_{\ell m,i}$ that are statistically independent in the wave numbers $\ell$ and $m$ despite incomplete sky coverage implying that the likelihood would not separate, but instead scale the logarithmic likelihood with a factor $\sqrt{f_\mathrm{sky}}$. The tomographic bins are chosen to contain the same fraction $1/n_\mathrm{bin}$ of galaxies.

\section{Variations of the Fisher matrix}\label{sect_lie}
Fisher-matrix forecasts depend on the choice of the fiducial model, because $(i)$ the model determines the signal strength relative to the noise and therefore the extent of the posterior likelihood and because $(ii)$ nonlinearities of the model are in general different at different points in parameter space, leading to varying degeneracies. In fact, it is apparent from the definition of the Fisher-matrix eqn.~(\ref{eqn_fisher}) that the true physical parameters are replaced by effective linear parameters, which in cosmology effectively applies to every parameter except $\sigma_8^2$ for linear structures: For $\sigma_8$, the derivative $\partial C_{\psi,ij}(\ell)/\partial\sigma_8^2 = \mathrm{const}$ in the regime of linear structure formation. Consequently, the Fisher-matrix is in general varying across parameter space, leading to changes in size and orientation of the posterior likelihood.

\foca{In particular in the context of weak gravitational lensing it is worth to discuss the effects of linear and nonlinear parameter dependences on one side and the two cases of cosmic variance domination and shape noise domination on the other. The Fisher-matrix is composed of terms $\partial_\mu\ln \hat{C}_{\psi,ij}$, where the covariance matrix $\hat{C}_{\psi,ij} = S_{ij}+N_{ij}$ contains the signal $S_{ij}$ as well as a noise term $N_{ij}$ which is independent of the model parameters. A linear model would be characterised by $S$ being proportional to the parameters $x_\mu$, in which case $\partial_\mu\ln C$ is constant if the noise is shape noise, $\partial_\mu\ln C = Q/N$ with $C=x_\mu Q$ in the case $S\ll N$. This is the case of a constant Fisher matrix with constant (absolute) errors $\sigma_\mu = (F^{-1})_{\mu\mu}$. In the case of cosmic variance domination, $N\ll S$ and $\partial_\mu\ln C$ becomes $1/x_\mu$: In this case the relative error $\sigma_\mu/x_\mu$ is constant and does not depend on the choice of the fiducial model, whereas the absolute error $\sigma_\mu$ does in fact vary across the parameter space: To some approximation, this is realised in weak lensing when considering the parameter $\sigma_8^2$. In nonlinear models, as in the case considered here, the covariance $C = Q g(x_\mu)$ depends on the model parameters through a nonlinear function $g(x_\mu)$, such that the derivative $\partial_\mu\ln C$ becomes $\partial_\mu\ln g$ in the case of cosmic variance domination, and $Q/N\times\partial_\mu g$ in shape noise domination, both results are not constant and lead again to variations of the Fisher-matrix. Whereas linear models will always imply Gaussian likelihoods and possibly constant Fisher-ellipses, the likelihood in the nonlinear case is only approximated by elliptic contours, which in addition, vary if a different fiducial cosmology is chosen, as a reflection of their nonlinear parameter dependence. At the same time, but not as a principal reason, the likelihood is naturally non-Gaussian and reduces to an approximately Gaussian shape if the data is very constraining, but nevertheless one observes variations in the curvature of the logarithmic likelihood, due to varying degeneracies.}

Another numerical effect of model nonlinearities are sometimes surprising dependencies of the Fisher matrix on the size of the parameter variation used in estimating the derivatives by finite differencing. This effect is completely absent in the case of linear parameters. Tomographic measurements with a high significance such as Euclid's weak lensing survey restrict the posterior distribution to a small region of parameter space in which a linear approximation of the model is applicable and where the posterior distribution therefore assumes a Gaussian shape. This has been established by comparison of Fisher-forecasts with MCMC-sampled likelihoods for the case of tomographic weak lensing measurements and cosmological models with the complexity of $w$CDM. Only weaker signals, or equivalently, models with more parameters force the likelihood outside the linear regime of the model, making the posterior distribution necessarily non-Gaussian. Therefore, we restrict ourselves to Gaussian posterior distributions in investigating the dependence of signal strength and parameter degeneracy on the choice of the fiducial model. A tool for dealing with non-Gaussian likelihoods on an analytical basis without resorting to MCMC is DALI \citep{Sellentinetal, 2015arXiv150605356S}, which uses a higher-order expansion of the logarithmic likelihood. 

\subsection{Construction of a Lie-basis}
We intend to formulate a basis for describing variations of Fisher matrices due to varying signal strength and model nonlinearities. This basis should relate Fisher-matrices throughout the parameter space to each other: Specifically, Fisher matrices  $F_{\mu\nu}(x^\prime_\alpha)$ and $F_{\rho\sigma}(x_\alpha)$ at two different fiducial models are related by a linear transform $U_{\mu\rho}$,
\begin{equation}
F_{\mu\nu}(x^\prime_\alpha) = U_{\mu\rho} F_{\rho\sigma}(x_\alpha) U_{\sigma\nu},
\end{equation}
where the transformation $U_{\mu\rho}$ will depend on the distance between the two points $x_\alpha$ and $x^\prime_\alpha$ in parameter space. Effectively, this means that the transformations $U_{\mu\rho}$ are continuously parameterised by the distance $x^\prime_\alpha-x_\alpha$ and form therefore a Lie-group. \spirou{Apart from linearity and orthogonality there is no restriction on the above transformation, in particular there might be a change in size, a rotation as well as a shearing effect of $U_{\mu\rho}$ on the Fisher-matrix $F_{\rho\sigma}$.}

The transformation can be constructed from a given pair of Fisher matrices by drawing the matrix root, $F_{\rho\sigma}(x_\alpha) = B_{\rho\tau}B_{\tau\sigma}$ and $F_{\mu\nu}(x^\prime_\alpha) = B^\prime_{\mu\sigma}B^\prime_{\sigma\nu}$ of each of the matrices and by identifying two identical pairs of terms of the type
\begin{equation}
B^\prime_{\mu\sigma} = U_{\mu\rho} B_{\rho\sigma}
\quad\rightarrow\quad 
U_{\mu\nu} = B^\prime_{\mu\rho} (B^{-1})_{\rho\nu},
\end{equation}
from which the matrix $U_{\mu\nu}$ can be isolated. If $x^\prime_\alpha$ approaches $x_\alpha$, the transformation will be to the unit matrix $\delta_{\mu\nu}$, so for small distances a linear approximation must apply,
\begin{equation}
U_{\mu\nu} \simeq \delta_{\mu\nu} + (x^\prime_\alpha-x_\alpha)T_{\mu\nu,\alpha},
\label{eqn_lin}
\end{equation}
which contains the generator $T_{\mu\nu,\alpha}$ of the transformation into the direction $x_\alpha$. It can be derived by differentiation,
\begin{equation}
T_{\mu\nu,\alpha} = 
\partial_\alpha U_{\mu\nu} =
\lim_{x_\alpha^\prime\rightarrow x_\alpha}\frac{U_{\mu\nu}(x_\alpha^\prime) - \delta_{\mu\nu}}{x_\alpha^\prime - x_\alpha}.
\end{equation}
Inversely, the transformation $U_{\mu\nu}$ can be constructed from the generator $T_{\mu\nu,\alpha}$ through the matrix exponential series as a sum of infinitesimal transformations,
\begin{equation}
U_{\mu\nu} = 
\exp\left((x^\prime_\alpha - x_\alpha) T_{\mu\nu,\alpha}\right),
\label{eqn_exp}
\end{equation}
which we will show to be sufficiently approximated by the first two terms of the matrix exponential. In this case, the approximated Fisher-matrix $A_{\mu\nu}(x^\prime_\alpha)$ would result from
\begin{equation}
A_{\mu\nu}(x^\prime_\alpha) = U_{\mu\rho} F_{\rho\sigma}(x_\alpha)U_{\rho\nu}
\quad\mathrm{with}\quad
U_{\mu\nu} \simeq \delta_{\mu\nu} + (x^\prime_\alpha-x^\prime) T_{\mu\nu,\alpha},
\end{equation}
which effectively amounts to a linear extrapolation of the Fisher-matrix variation from the fiducial model at $x_\alpha$ to the new model $x^\prime_\alpha$. In fact, assembling the transformation $U_{\mu\nu}$ from the generators $T_{\mu\nu,\alpha}$ yields an abelian group of transformations either in the case of linear extrapolations or if the generators describe only size changes, or if there is only a single rotation or shearing distortion involved. This is a necessary condition because of the uniqueness of the transformation irrespective of the order of summation in the parameter space directions $\alpha$.

Numerically, we derive $T_{\mu\nu,\alpha}$ by finite differencing from the transformation $U_{\mu\nu}$, which in turn has been derived from Fisher-matrices at two neighbouring points $x_\alpha\pm\Delta x_\alpha$,
\begin{equation}
T_{\mu\nu,\alpha} \simeq 
\frac{1}{2\Delta x_\alpha}
\left[U_{\mu\nu}(x_\alpha+\Delta x_\alpha) - U_{\mu\nu}(x_\alpha-\Delta x_\alpha)\right],
\end{equation}
with a typical variation $\Delta_\alpha$ amounting to a few percent of the fiducial value of $x_\alpha$. The linear approximation is sufficient given the slow variation of $F_{\mu\nu}$, as will be shown by Figs.~\ref{fig_os},~\ref{fig_ow} and~\ref{fig_ww}.

An alternative description of Fisher-matrix variations is due to \citet{santi}, who perform a Taylor-expansion of the Gaussian likelihood, $F_{\mu\nu}(x_\alpha+\Delta x_\alpha) \simeq \partial_\alpha F_{\mu\nu}(x_\alpha)\:\Delta x_\alpha$ to first order, and find that it captures variations well, commensurate with our experience that variations across parameter space are weak. In order to carry out marginalisations we think that our method might be slightly simpler, as a sum of matrices is not straightforward to invert, although approximations exist if the term $\partial_\alpha F_{\mu\nu}\:\Delta x_\alpha$ is small.

Conditionalisations are, in contrast, straightforward in both formalisms but for combining different Gaussian likelihoods or priors, the direct Taylor-expansion is better suited, due to the additivity of the Fisher-matrix for independent likelihoods: In this case, there is no direct way, due to the matrix root in constructing the transformation, to combine generators. Instead one needs to compute Fisher-matrices for each likelihood separately and add them in a second step. The respective advantages and disadvantages originate from describing the variation of the Fisher-matrix with a multiplication in our case or by an addition in the case of \citet{santi}.

If one chooses a reparameterisation $y_\beta(x_\alpha)$, for instance by applying a principal component analysis which in the Gaussian case decouples otherwise correlated parameters, it is straightforward to show that the generators transform according to the rule
\begin{equation}
T_{\mu\nu,\alpha} \rightarrow J_{\mu\rho} J_{\nu\sigma} J^{-1}_{\alpha\beta}T_{\rho\sigma,\beta}
\end{equation}
with the Jacobian matrices relating the old parameterisation $x_\alpha$ to a new one $y_\beta$ through $J_{\alpha\beta} = \partial x_\alpha/\partial y_\beta$. In this way, the transformation $U_{\mu\nu}$ changes, when generated from $T_{\mu\nu,\alpha}$, in analogy to the Fisher-matrix $F_{\mu\nu}$, $F_{\mu\nu}\rightarrow J_{\mu\alpha}J_{\nu\beta}F_{\alpha\beta}$, which itself is a tensor.

\subsection{Distortion modes}
In two dimensions one can apply a decomposition of the transformation $U_{\mu\nu}$ into the set of Pauli-matrices $\sigma^{(n)}_{\mu\nu}$ with the unit matrix for completeness, $\sigma^{(0)}_{\mu\nu}=\delta_{\mu\nu}$,
\begin{equation}
U_{\mu\nu} = 
\sum_{n=0}^3 a_n\sigma^{(n)}_{\mu\nu} = 
(1+\kappa)\sigma^{(0)}_{\mu\nu} + \gamma_+\sigma^{(1)}_{\mu\nu} + \omega\sigma^{(2)}_{\mu\nu} + \gamma_\times\sigma^{(3)}_{\mu\nu},
\end{equation}
where the interpretation of the coefficients is in complete analogy to gravitational lensing: $\kappa$ describes an overall isotropic change in size of the Fisher matrix, $\omega$ a rotation, and $\gamma_+$ and $\gamma_\times$ jointly shearing. As in gravitational lensing, we combine the two components of shear to a single parameter, $\gamma^2 = \gamma_+^2+\gamma_\times^2$. These coefficients can easily be obtained by inversion, $a_n = U_{\mu\nu}\sigma^{(n)}_{\nu\mu}/2$, due to the relation $\sigma^{(a)}_{\mu\sigma}\sigma^{(b)}_{\sigma\nu} = \delta_{ab}\sigma^{(0)}_{\mu\nu}+\ci\epsilon_{abc}\sigma^{(c)}_{\mu\nu}$ and because of the trace $\sigma^{(0)}_{\mu\mu}=2$.

In this way, one can quantify easily specific changes of the Fisher-matrix across parameter space. Additionally, computing the change of the dark energy figure of merit $\det(F_{\mu\nu})$ with the cosmological parameters is simply given by the factor $\det(U_{\mu\sigma}U_{\sigma\nu})$. In fact, only for the case of $U_{\mu\nu}$ consisting of a pure rotation, $U_{\mu\nu}\propto\sigma^{(2)}_{\mu\nu}$ with a rotation angle $\omega$, the transformation will be orthogonal and conserve e.g. the dark energy figure of merit as the determinant $\det(U_{\mu\sigma}U_{\sigma\mu})$ is equal to unity.

Alternatively, it would be equally possible to decompose the generators $T_{\mu\nu,\alpha}$ for each direction $\alpha$. There, similar arguments apply to commutativity and orthogonality, in particular in the linear approximation eqn.~(\ref{eqn_lin}) and the full expression eqn.~(\ref{eqn_exp}): Commutativity of $\sigma^{(n)}_{\mu\nu}$ implies commutativity of the corresponding transformations $U_{\mu\nu}$, but naturally, only the unit matrix commutes with the other matrices of the Pauli-basis, for which there is the relation $[\sigma^{(i)},\sigma^{(j)}] = 2\epsilon_{ijk}\sigma^{(k)}$. Commutativity of the transformations $U_{\mu\nu}$ is then a consequence of the Baker-Hausdorff-identity,
\begin{equation}
\exp\left(T_{\mu\nu,\alpha} + T_{\mu\nu,\beta}\right) = 
\exp\left(T_{\mu\nu,\alpha}\right) \exp\left(T_{\mu\nu,\beta}\right) \exp\left(\left[T_{\mu\nu,\alpha},T_{\mu\nu,\beta}\right]/2\right),
\end{equation}
which illustrates commutativity of the set $U_{\mu\nu}=\exp(T_{\mu\nu,\alpha})$ for different directions $\alpha, \beta$ if the commutator $\left[T_{\mu\nu,\alpha},T_{\mu\nu,\beta}\right]$ vanishes. For simplicity, we have set the shift $x^\prime_\alpha-x_\alpha$ to unity. Any transformation involving the unit matrix is commutative, but not the general case: This is circumvented by considering only linear approximations, which are always commutative at the order $(T_{\mu\nu,\alpha})^2$.

Orthogonal transformations $U_{\mu\nu}$ only result from the rotation $\sigma^{(2)}_{\mu\nu}$ because of its antisymmetry, $\sigma^{(2)}_{\mu\nu} = -\sigma^{(2)}_{\mu\nu}$, because $U_{\mu\nu}U_{\nu\mu} = \exp(\sigma^{(2)}_{\mu\nu} + \sigma^{(2)}_{\nu\mu}))=\exp(0\times\sigma^{0}_{\mu\nu}) = \delta_{\mu\nu}$. Naturally, rotations conserve the determinant of $F_{\mu\nu}$ and therefore the dark energy figure of merit.

\subsection{Application to weak lensing forecasts}
As an application of the formalism we consider Euclid's weak lensing forecasts on three sections in parameter space: the $(\Omega_m,\sigma_8)$-plane, the $(\Omega_m,w_0)$-plane and finally the $(w_0,w_a)$-plane.

\begin{figure*}
\begin{center}
\resizebox{0.98\hsize}{!}{\includegraphics{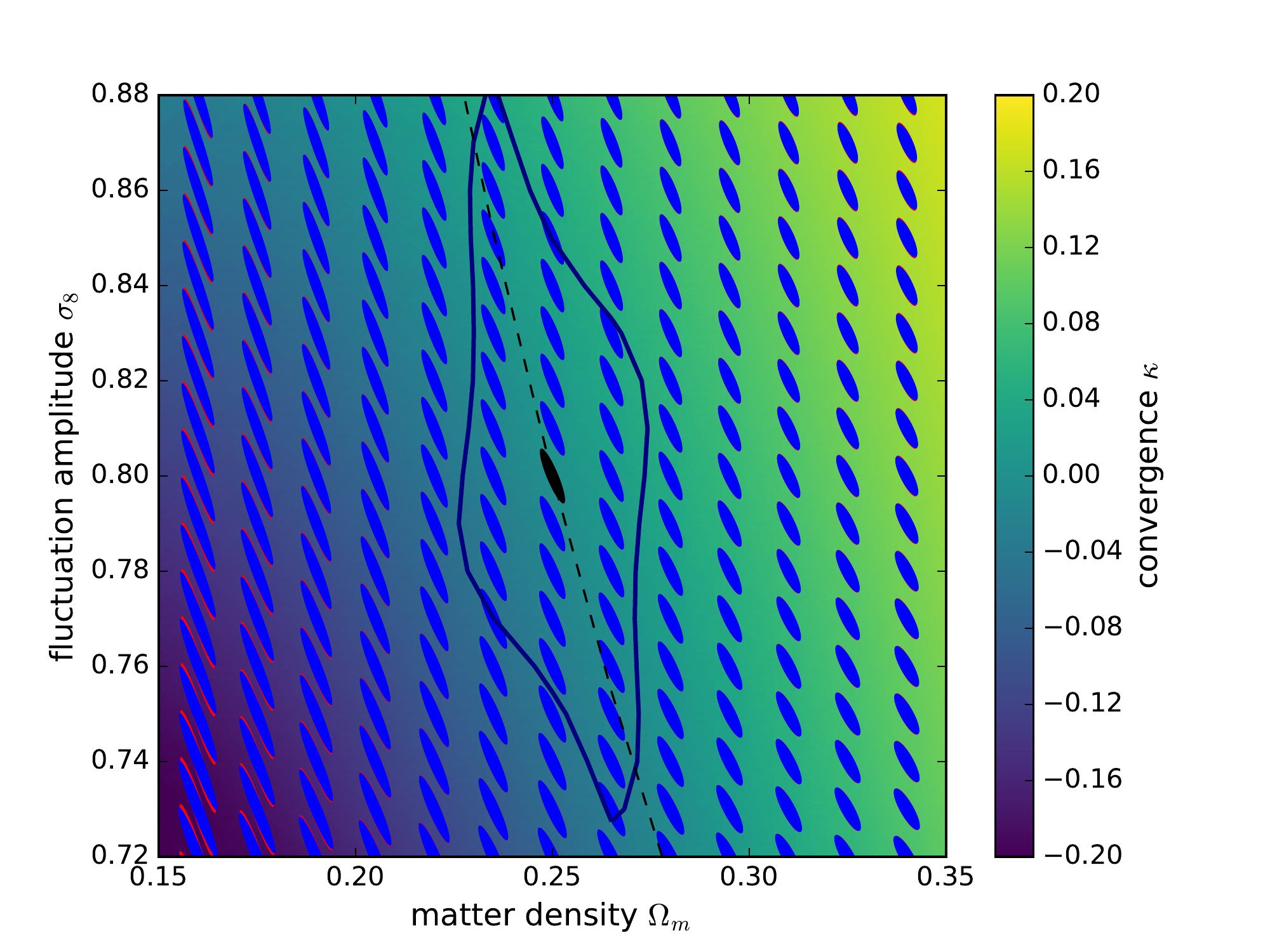}}
\end{center}
\caption{Fisher ellipses in the $\Omega_m$-$\sigma_8$-plane (red, in the background), along with the Lie-approximated ellipses (blue, in the foreground). The $\Lambda$CDM-reference ellipse is drawn in black at the centre of the plot. The background colour indicates the (isotropic) change in size of the Fisher-ellipse relative to the fiducial due to a different choice of parameters, and the black contour shows the zone in which the Fisher-ellipses can be predicted with an error of less than 0.005. The sky coverage was taken to by $f_\mathrm{sky}=0.5$. Cosmologies with a constant product $\Omega_m\times\sigma_8$ lie on the black dashed line.}
\label{fig_os}
\end{figure*}

The magnitude of the lensing signal is primarily set by the two cosmological parameters $\Omega_m$ and $\sigma_8$, which determine the strength of the gravitational potential: This well-known fact leads to a banana-shaped likelihood in the $(\Omega_m,\sigma_8)$-plane which has been observed by e.g. CFHTLenS \citep{kilbinger_dark-energy_2009,kitching_3d_2014}. Consequently, one would expect a change in the size in the extent of the posterior distribution between large products $(\Omega_m\sigma_8)^2$, which generate strong lensing signals and therefore small statistical errors and small products $(\Omega_m\sigma_8)^2$. This behaviour is recovered by Fig.~\ref{fig_os}, either directly in the size of the $1\sigma$-ellipses or in the background color gradient, which depicts the magnitude of the size change through the decomposition of the matrix $U$, through $U_{\mu\nu}\sigma^{(0)}_{\nu\mu}$. Additionally, there is effectively no change in the extent of the posterior distribution in models that keep the product $(\Omega_m\sigma_8)^2$ constant.

A comparison between the true Fisher-ellipse $F_{\mu\nu}(x^\prime_\alpha)$ and $A_{\mu\nu}(x^\prime_\alpha)$ predicted by the Lie-generator at the position $x^\prime_\alpha$ at linear order yields a very good agreement, even when varying the cosmological parameters by a large amount. Among many possibilities we quantify the difference between the true and the approximated Fisher-matrix by evaluating a relative difference between the two matrices,
\begin{equation}
\frac{\|F_{\mu\nu}(x^\prime_\alpha)-A_{\mu\nu}(x^\prime_\alpha)\|}{\|F_{\mu\nu}(x^\prime_\alpha)\|},
\label{eqn_frobenius}
\end{equation}
with the Frobenius matrix norm
\begin{equation}
\| A_{\mu\nu} \| = \sqrt{A_{\mu\nu}A_{\nu\mu}}.
\end{equation}
In fact, it is possible to predict the Fisher matrices accurate to better than 0.005 for variations of $\Omega_m$ and $\sigma_8$ of up to 10\% of their fiducial values, and this region is aligned with the primary degeneracy direction along constant ($\Omega_m\sigma_8)^2$.

\begin{figure*}
\begin{center}
\resizebox{0.98\hsize}{!}{\includegraphics{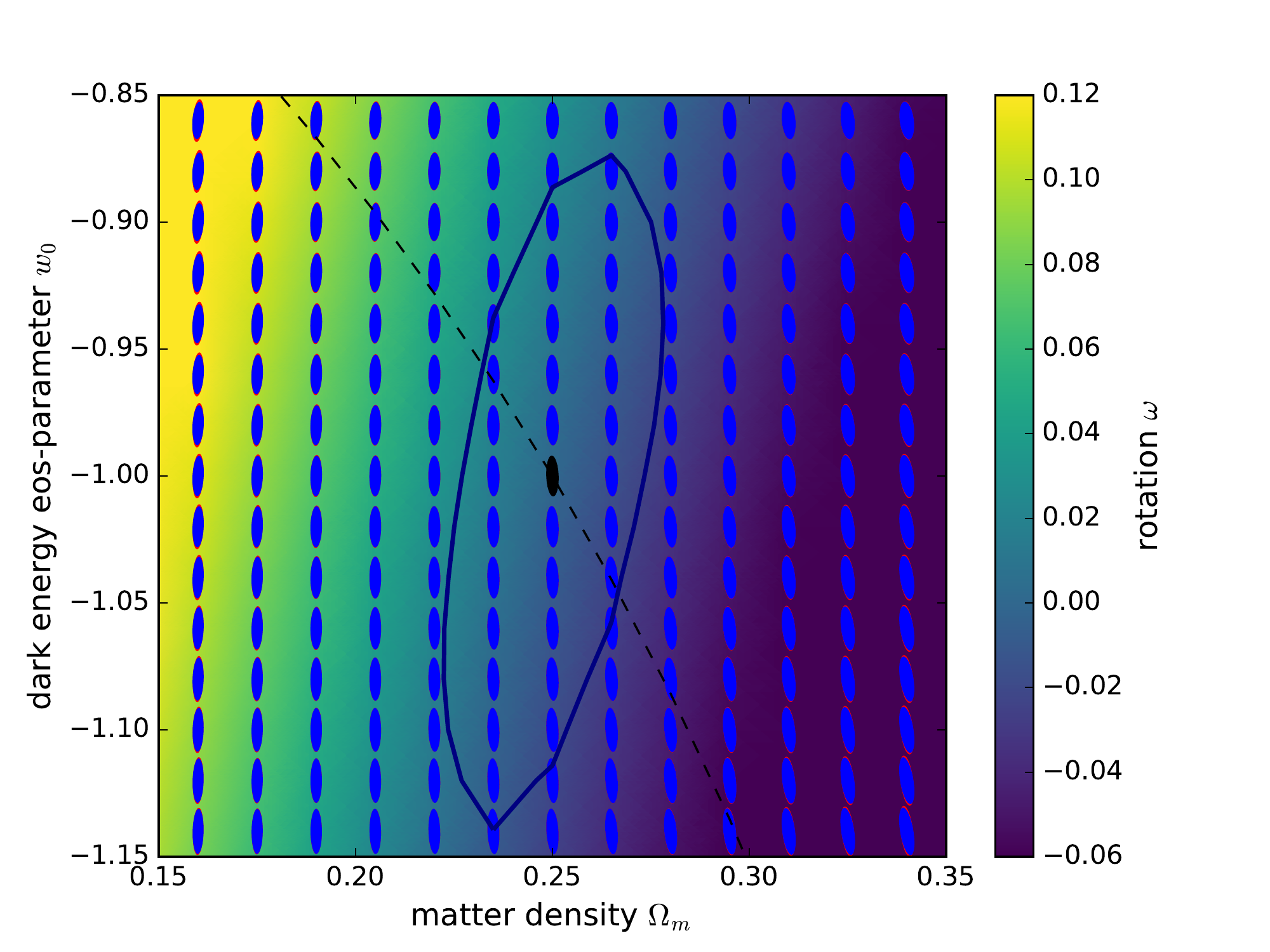}}
\end{center}
\caption{Fisher ellipses (red, in the background) and their Lie-approximations (blue, in the foreground) in the $\Omega_m$-$w_0$-plane. The background colour shows the rotation of the Fisher-ellipses relative to the fiducial model, and the black contour indicates the zone where the Lie-approximation deviates from the exact Fisher-matrix by less than 0.005. The ellipses assume a sky coverage of $f_\mathrm{sky}=0.25$. The black dashed lines marks cosmologies with equal Hubble-parameters at $a=2/3$. The Fisher ellipse corresponding to the fiducial cosmology is drawn in black, at the centre of the plot.}
\label{fig_ow}
\end{figure*}

Similar results apply to the $(\Omega_m,w_0)$-plane: There is a smooth rotation of the Fisher-ellipses with increasing matter density $\Omega_m$, while the size of the ellipses stays essentially constant, as shown by Fig.~\ref{fig_ow}. The true Fisher-ellipses can be predicted to an accuracy of 0.005 with the generators in a region in which $\Omega_m$ varies by about 20\% and where $w_0$ varies by 10\% around their fiducial values, and this region coincides with that of similar orientation of the ellipses, indicating that the rotation is in fact the dominant distortion mode. Physically, larger densities of dark energy, or smaller values of $\Omega_m$ lead to an enhanced growth $D_+/a$ of the potentials if the growth function is normalised to one today, $D_+(1)=1$, likewise a less negative equation of state parameter $w$. At the same time, the Hubble function $H(a)$ which converts the redshift distribution $p(z)\dd z$ into a distribution of comoving distance, assumes the smallest values for both small $\Omega_
m$ and more negative $w$, if its value today is fixed, $H(1)=H_0$. Additionally, the lensing signal is sensitive to $\Omega_m$ through the shape parameter $\Gamma\simeq\Omega_m h$, which changes the amplitude of the CDM-spectrum $P(k)$, which has a minor effect on the lensing spectrum.

Products of strong growth and small Hubble functions should cause degeneracies in the lensing signal, i.e. combinations of small $\Omega_m$ and less negative $w$ for growth and small $\Omega_m$ with more negative $w$ for the Hubble functions, leading to almost independent measurements. The line of equal Hubble functions and the line of equal growth rates would run through the parameter plane diagonally. The lensing signal constrains $\Omega_m$ tighter than $w$, due to the fact that weak cosmic shear strength is proportional to $\Omega_m$ but measures $w$ only through an integral over the scale factor $a$.

\begin{figure*}
\resizebox{0.98\hsize}{!}{\includegraphics{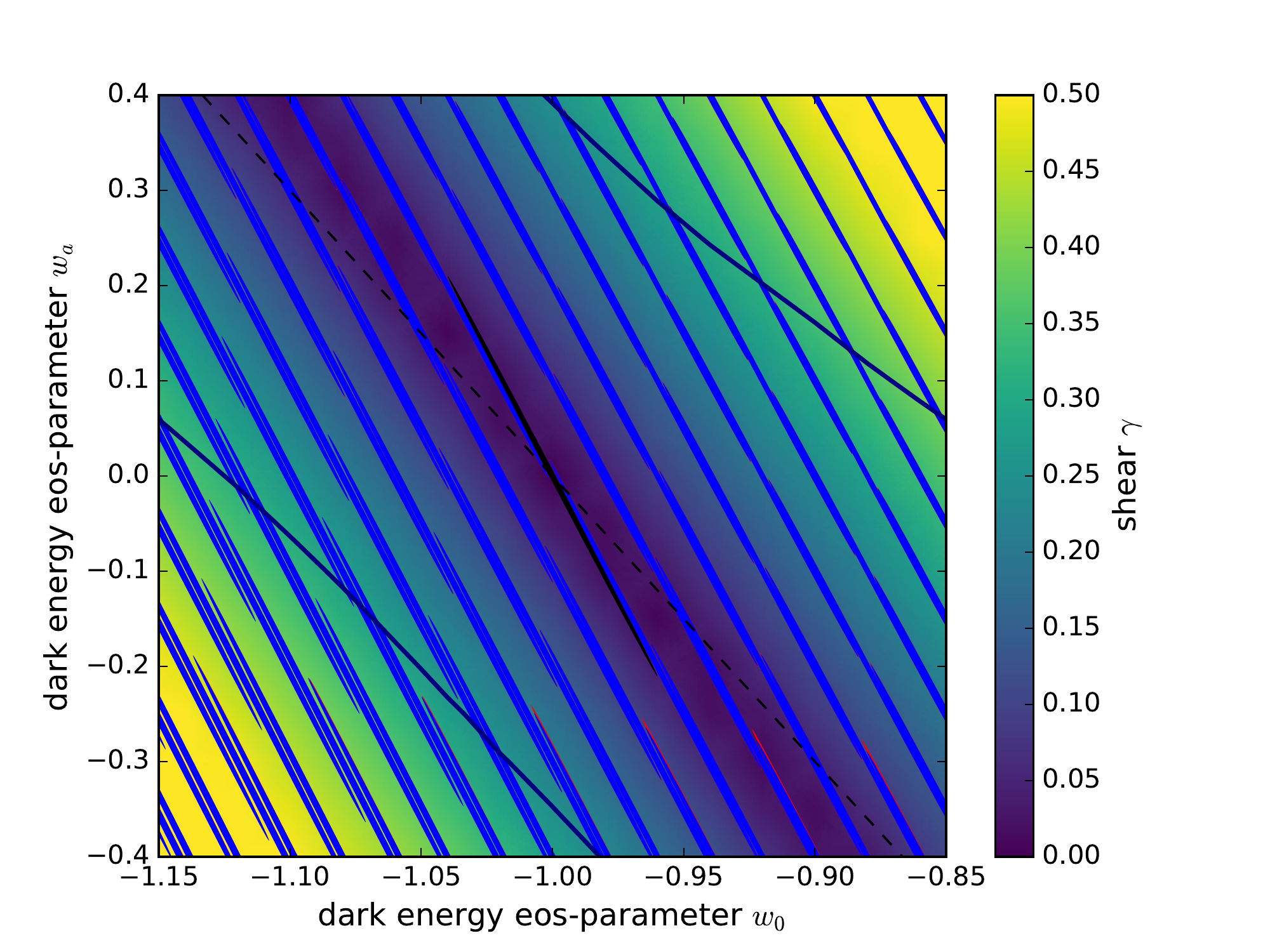}}
\caption{Fisher constraints in the $w_0$-$w_a$-plane (red, in the background) and the corresponding Lie-approximations (blue, in the foreground). Inside the black contour it is possible to predict the Fisher ellipses with an accuracy better than 0.005. The ellipses have been scaled down with a factor of 0.1 due to the strong degeneracy between $w_0$ and $w_a$. The dashed black line indicates cosmologies with $w(a)=-1$ at $a=2/3$, and the Fisher-ellipse of the fiducial cosmology is drawn in black at the centre of the plot.}
\label{fig_ww}
\end{figure*}

Lastly, the $(w_0,w_a)$-plane shows large degeneracy between the two dark energy parameters which is due to the fact that the equation of state function $w(a)$ enters lensing observables through an integral over scale factor $a$, and through the integration process in solving the growth equation, before one carries out the line of sight integration as the third integral. Therefore, weak lensing alone is only sensitive to an effective equation of state, causing large degeneracies between $w_0$ and $w_a$, which can only partially be broken by lensing tomography, as depicted by Fig.~\ref{fig_ww}. Apart from being interesting in its own right, we investigated Fisher-ellipses in that particular plane in parameter space to show that the formalism is able to deal with strongly elongated ellipses. In fact, one observes a shearing (along with a rotation) in the Fisher-ellipses when varying the two dark energy parameters, and a wide region in which the ellipses can be predicted accurately which is aligned with the direction of constant effective equations of state and therefore with the primary degeneracy direction of the Fisher-ellipses. 

\spirou{There is a dependence of the precision at which dark energy equation of state parameters can be measured: Computing the Fisher-matrix $F_{w_0,w_a}$ in the $w_0$-$w_a$-plane shows that models with very negative effective equations of state can be better constrained than those with less negative ones.} This can be confirmed in two ways, either by computing the trace $U_{\mu\mu} = U_{\mu\sigma}\sigma^{(0)}_{\mu\nu}$ or the determinant $\det(U_{\mu\sigma}U_{\sigma\mu})$. Both are indicators of the change in area of the ellipse in the $w_0$-$w_a$-plane, which we found to be slowly varying across this parameter space section, with typical  variations between $-26\%$ and $+30\%$ relative to the area at the fiducial $\Lambda$CDM-cosmology. Again, variations of the Fisher-matrix can be reliably described with our formalism: Within a 10\%-variation in $w_0$ \citep[which roughly corresponds to the current uncertainty on this parameter,][]{planck_collaboration_planck_2015} the relative extrapolation error amounts to 0.005.

\begin{figure}
\resizebox{0.98\hsize}{!}{\includegraphics{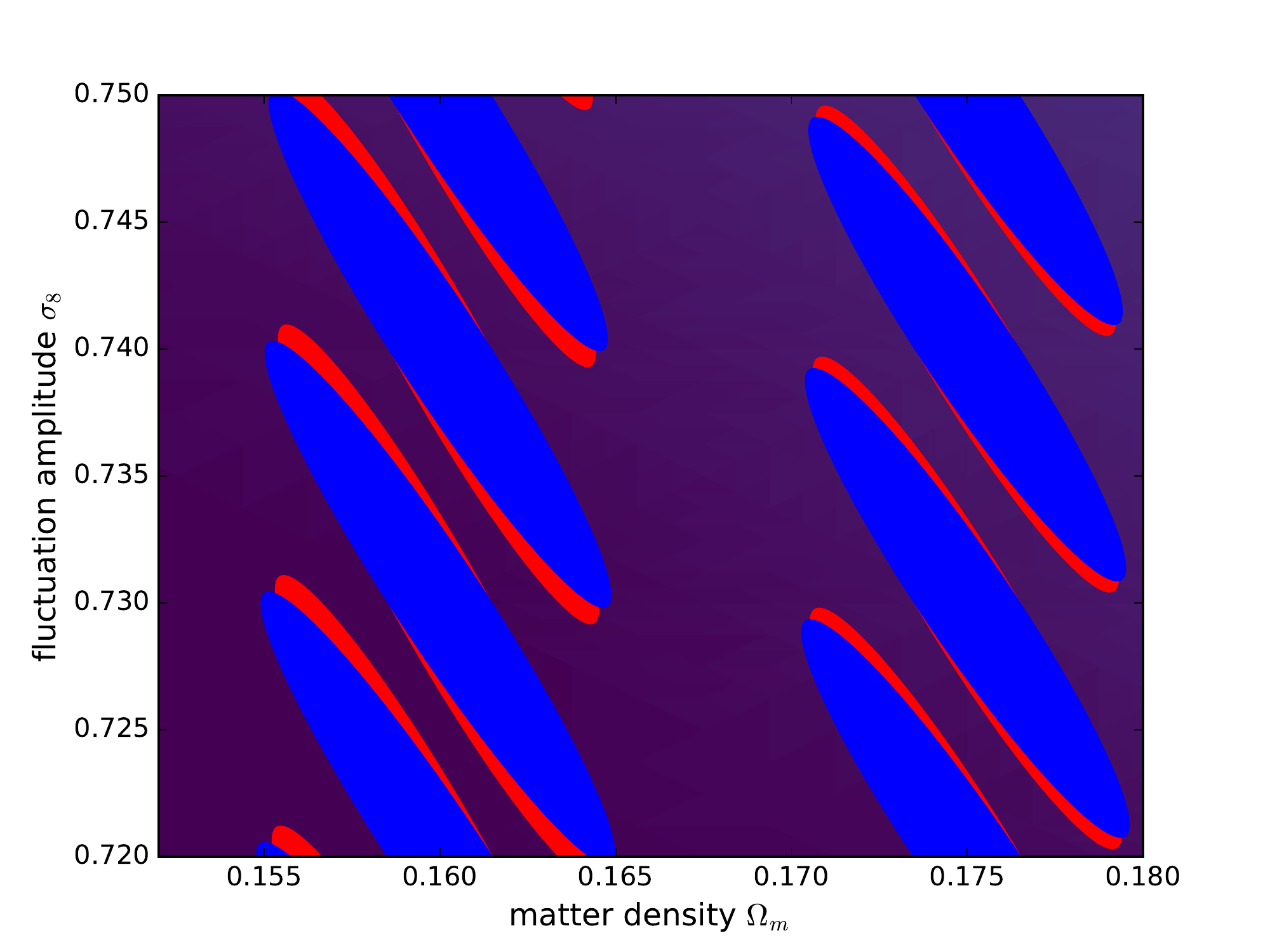}}
\caption{Fisher constraints in the $\Omega_m$-$\sigma_8$-plane (red, in the background) and the corresponding Lie-approximations (blue, in the foreground) in the (low-$\Omega_m$, low-$\sigma_8$)-corner of the parameter space where the deviations between the approximated and the true Fisher-matrix are strongest. The difference between the two ellipses, quantified in terms of the Frobenius-norm (c.f. eqn.~\ref{eqn_frobenius}) amounts to about 1\%.}
\label{fig_zoom}
\end{figure}

\spirou{In summary it is welcome that there are only comparatively small variations of the forecasts across parameter space, both in size, ellipticity and orientation of the Fisher-ellipses, at least for a $w$CDM cosmology. This is exemplified in Fig.~\ref{fig_zoom}, where we plot a zoom into a corner of the $\Omega_m$-$\sigma_8$-plane quite far from the fiducial cosmology and show the true and the extrapolated Fisher-ellipses in comparison.}

Lastly, we would like to add a few remarks on the treatment of systematical errors in our formalism, as future surveys will be very likely dominated by systematics instead of statistics. Formalisms for predicting biases in the estimation of cosmological parameters have been developed in a number of different applications \citep[in the context of CMB-science,][]{cabre_error_2007, taburet_biases_2009}, and generalised to the case of non-vanishing covariances between data points \citep{amara_systematic_2008, 2012MNRAS.423.3445S}. In this formalism, the bias $\delta_\mu$ results from solving a linear system $G_{\mu\nu}\delta_\mu = a_\nu$ where the matrix $G_{\mu\nu}$ and the vector $a_\nu$ are computed from the signal, its noise and the systematic in question. Both quantities change across parameter space, and one can derive transformations for $G_{\mu\nu}$ and $a_\nu$ separately, which can be used to compute the variation of $\delta_\mu$, and in this way any complication arising from finding a basis for $G^{-1}_{\mu\nu}$ is avoided. Specifically, the basis for $G_{\mu\nu}$ is constructed in complete analogy to that of $F_{\mu\nu}$, whereas only a single transformation matrix is needed for the vector $a_\mu$.

Finally we would like to point out that generators $T_{\mu\nu,\alpha}$ can be used to predict Fisher-matrices and derived quantities such as the dark energy figure of merit with the parameters characterising survey design \citep{kirk_optimising_2011}, in order to carry out optimisations: In this case, $T_{\mu\nu,\alpha}$ would describe the rate of change of the Fisher-matrix with a survey parameter such as redshift depth or shape noise amplitudes.

\section{Summary}\label{sect_summary}
Subject of this paper was the investigation of the model parameter dependence of statistical forecasts made with the Fisher-formalism, specifically tomographic weak lensing forecasts for Euclid, and a description of this parameter dependence with a specifically constructed basis. In general, we found that in the three sections through a $w$CDM-parameter space there was a weak variation of the Fisher-constraints with the choice of the fiducial cosmology, in particular the forecasted precision on the dark energy equation of state is almost independent from the fiducial.

\begin{enumerate}
\item{From the Fisher-matrix at some position in parameter space and the Fisher-matrix at the reference cosmology we derived a linear transformation that relates the two. Expanding the transformation matrix in terms of Pauli-matrices leads to a straightforward interpretation of the variations of the Fisher-matrices in terms of changes in size, rotation and shearing.}
\item{We constructed a basis from the transformation matrix between the Fisher-matrices at the fiducial cosmology and an infinitesimally displaced model. Transformations for finite displacements are derived by linear extrapolation. Comparing the extrapolated Fisher-matrix with the true one shows an accurate prediction for a large region in parameter space, which coincides as expected with the main degeneracy directions.}
\item{Specifically, the $(\Omega_m,\sigma_8)$-plane is known for having strong degeneracies in gravitational lensing, ultimately a hyperbolic banana-shaped confidence contour for weaker signals. There, the variation of the Fisher-ellipse was predicted to percent accuracy for variations of 10\% in $\Omega_m$ and $\sigma_8$. At the same time, the main change is simply a scaling in size of the Fisher-matrix, reflecting the proportionality of the lensing signal to the product $\propto(\Omega_m\sigma_8)^2$. Likelihood ellipses in the ($\Omega_m,w_0)$-plane show a rotation with $\Omega_m$, while staying roughly constant in size, and while showing a similarly large range of accurate extrapolations. In all cases, directions of weak changes in parameter space can be understood as weak dependences of the weak lensing model on parameters, such as lines of constant $(\Omega_m\sigma_8)^2$ or constant average dark energy equation of state parameters.}
\item{The formalism is capable of dealing with strongly degenerate cases such as the $(w_0,w_a)$-constraints from weak lensing alone, which shows a shearing across parameter space: There, the variation of the precision at which dark energy parameters can be measured is accurately predicted for the currently allowed parameter space.}
\item{The method is limited by the volume of parameter space where linear extrapolation is applicable: In general, commutativity of individual transforms along single parameters is not ensured. In one or two dimensions it should be possible to follow variations of the Fisher-matrix further, but errors will increase on a scale which can be estimated by the inverse of the generator, $\left|x^\prime_\alpha-x_\alpha\right|\simeq 1/\left|T_{\mu\nu,\alpha}\right|$.}
\item{We considered variations of Gaussian likelihoods, which are completely described by positive definite covariance matrices. Generalising this method to non-Gaussian likelihoods should be possible by considering the transformation of each term of an expansion of the logarithmic likelihood in polynomials, as in the DALI-formalism: This will be a set of generators $T_{\mu\nu\rho,\alpha}$ and $T_{\mu\nu\rho\sigma,\alpha}$ for the fourth order expansion, with $n^2(n+1)/2$ and $n^2(n+1)^2/4$ values for $n$ parameters, respectively.}
\end{enumerate}

Future developments include an extension of this formalism to estimation biases, which depend through both the Fisher-matrix and the dependency of any systematic on the set of cosmological parameters. The formalism above should be able to describe variations of a multivariate Gaussian in the the proposal distribution of an MCMC-engine, which could potentially speed up MCMC-sampling by reorienting and resizing the proposal distribution. As a third application, we envision studies of combining probes with large intrinsic degeneracies to fit cosmological data, as shown by \citet{2015arXiv150704351L} for signs for interacting dark matter based on CMB-data, or for studies on degeneracy breaking \citep{2012JCAP...04..027H, 2014PhRvD..89h3501A}.

The formalism is as well suited for describing variations of covariances of power spectrum estimates, which become non-diagonal in nonlinear structure formation, and is capable of interpolating between estimates of the covariance from simulations \citep{2015MNRAS.446.1756B}, or estimates of the data covariance matrix, which is needed for the precision determination of cosmological parameters \citep{kayo_information_2013, 2013MNRAS.432.1928T, 2014MNRAS.442.2728T}. We intend to investigate these questions further, as well as the application to systematical errors.

\section*{Acknowledgements}
We thank Luca Amendola, Valeria Pettorino, Elena Sellentin, Santiago Casas and Alina Kiessling for helpful discussions and support, and the referee, whose careful remarks helped us to clarify the paper. RR acknowledges funding through the graduate college {\em Astrophysics of cosmological probes of gravity} by Landesgraduiertenakademie Baden-W{\"u}rttemberg.

\bibliographystyle{mnras}
\bibliography{references}

\bsp
\label{lastpage}
\end{document}